\documentclass[a4paper,fleqn,usenatbib]{mnras}

\usepackage{newtxtext,newtxmath}

\usepackage[T1]{fontenc}
\usepackage{ae,aecompl}

\usepackage{graphicx}	
\usepackage{amsmath}	
\usepackage{amssymb}	

\usepackage{hyperref}
\usepackage{siunitx}

\title[Optical Polarimetry of KIC 8462852 in May--Aug 2017]{Optical Polarimetry of KIC 8462852 in May--August 2017} 

\author[I. A. Steele et al.]{
I. A. Steele,$^1$,
C. M. Copperwheat,$^1$
H. E. Jermak,$^1$
G. M. Kennedy,$^2$ and
G. P. Lamb$^1$ \\
$^{1}$Astrophysics Research Institute, Liverpool John Moores University, 146 Brownlow Hill, L3 5RF, UK\\
$^{2}$Department of Physics, University of Warwick, Gibbet Hill Road, Coventry, CV4 7AL, UK
}

\date{Accepted 2017 September 18. Received 2017 September 18; in original form 2017 September 7}

\pubyear{2017}

\begin{document}
\label{firstpage}
\pagerange{\pageref{firstpage}--\pageref{lastpage}}
\maketitle

\begin{abstract}
We present optical polarimetry in the period May--August 2017 of the enigmatic ``dipping'' star KIC 8462852.  During that period three $\sim1$\% photometric dips were reported by other observers. We measured the average absolute polarization of the source, and find no excess or unusual polarization compared to a nearby comparison star.  We place tight upper limits on any change in the degree of polarization of the source between epochs in- and out-of-dip of $<0.1$\% (8500-{\AA}) and $<0.2$\% (7050-{\AA} and 5300-{\AA}).   
How our limits are interpreted depends on the specific model being considered.  If the whole stellar disk were covered by material with an optical depth of $\sim$0.01 then the fractional polarisation introduced by this material must be less than 10-20\%. While our non-detection does not constrain the comet scenario, it predicts that even modest amounts of dust that have properties similar to Solar System comets may be detectable.  We note that the sensitivity of our method scales with the depth of the dip.  Should a future $\sim 20$\% photometric dip be observed (as was previously detected by {\em Kepler}) our method would constrain any induced polarization associated with any occulting material to 0.5--1.0\%.
\end{abstract}




\section{Introduction}
\label{sec:intro}  

\cite{boyajian} presented a {\em Kepler} lightcurve of
KIC 8462852 (TYC-3162-665-1, Boyajian's Star) which showed a number of irregularly shaped, aperiodic dips
in flux of up to 20 per cent and duration 5- to 80-d from 2009 May until the end of monitoring in 2013 May.  Over the same time period the source appeared to fade by $\sim3$\% \citep{montet}.  \citet{boyajian} also presented spectroscopy which showed the source is an apparently normal F3V star and shows no radial velocity variations.  
They discussed various possible interpretations of the dips.  Intrinsic variability is
ruled out since the detailed light curve behaviour 
and spectral type of the source are not consistent with any known variable 
source that shows aperiodic dips (principally R Coronae Borealis and Be Stars).
The most likely scenario is therefore some form of extrinsic variability,
i.e. occultation by circumstellar or interstellar material.  UX Ori and AA Tau 
systems show similar dipping optical lightcurves.  However these are young
stars with a strong infrared excess which is not present in KIC 8462852 \citep{lisse,meng}.
More generally the large dip amplitudes indicate that the occulting source is unlikely to be a solid body and is most likely associated with some form of clumpy/dusty material in either
a distant, optically thin region or part of multiple isolated objects such as (possibly broken up) comets or planetesimals \citep{etangs,boyajian}.  The relatively small
wavelength dependence of the long term dimming (compared to the galactic extinction law) reported by \cite{meng} indicates any dusty material is likely to be circumstellar rather than interstellar.

\cite{families} showed  that the timing of the {\em Kepler} dips was consistent with them being randomly distributed.  They investigated a family of plausible explanations and concluded that explanations associated with structure in the interstellar medium or an intervening disk were most likely.  On the other hand \cite{bodman} investigated the transit of a large comet family in detail.  They found that a single comet of similar size to those in our solar system produces 
a transit depth of the order of $10^{-3}$ lasting less than a day  which is much smaller and shorter than observed.  
However a large ($\sim100$) cluster of comets could fit the observed depths and durations.  

\cite{thompson} presented millimetre and 
submillimetre (SCUBA-2) continuum observations ``out of dip''. No significant emission is detected. They argue their low limits make a catastrophic planetary disruption hypothesis unlikely.  Alternatively \cite{metz} argue the historic consumption of a planet can be used to explain the possible long term dimming and lack of infrared excess with planet debris or out-gassing of tidally detached moons explaining the transient dips.  However the apparent brightening episodes found in long term ASAS data \citep{asas} may argue against this interpretation.

Dust scattering is a well known
source of polarization.    Given the significant depth of
the dips, it may be that a detectable polarization signal is expected if a comet or other dust related explanation is correct.  KIC 8462852 displayed its first detected dips since the end of {\em Kepler} monitoring in 2017 May--Aug.  Here we present the results of  approximately nightly polarimetric monitoring over that period. 

   \begin{figure*}
   \includegraphics[width=16.0cm]{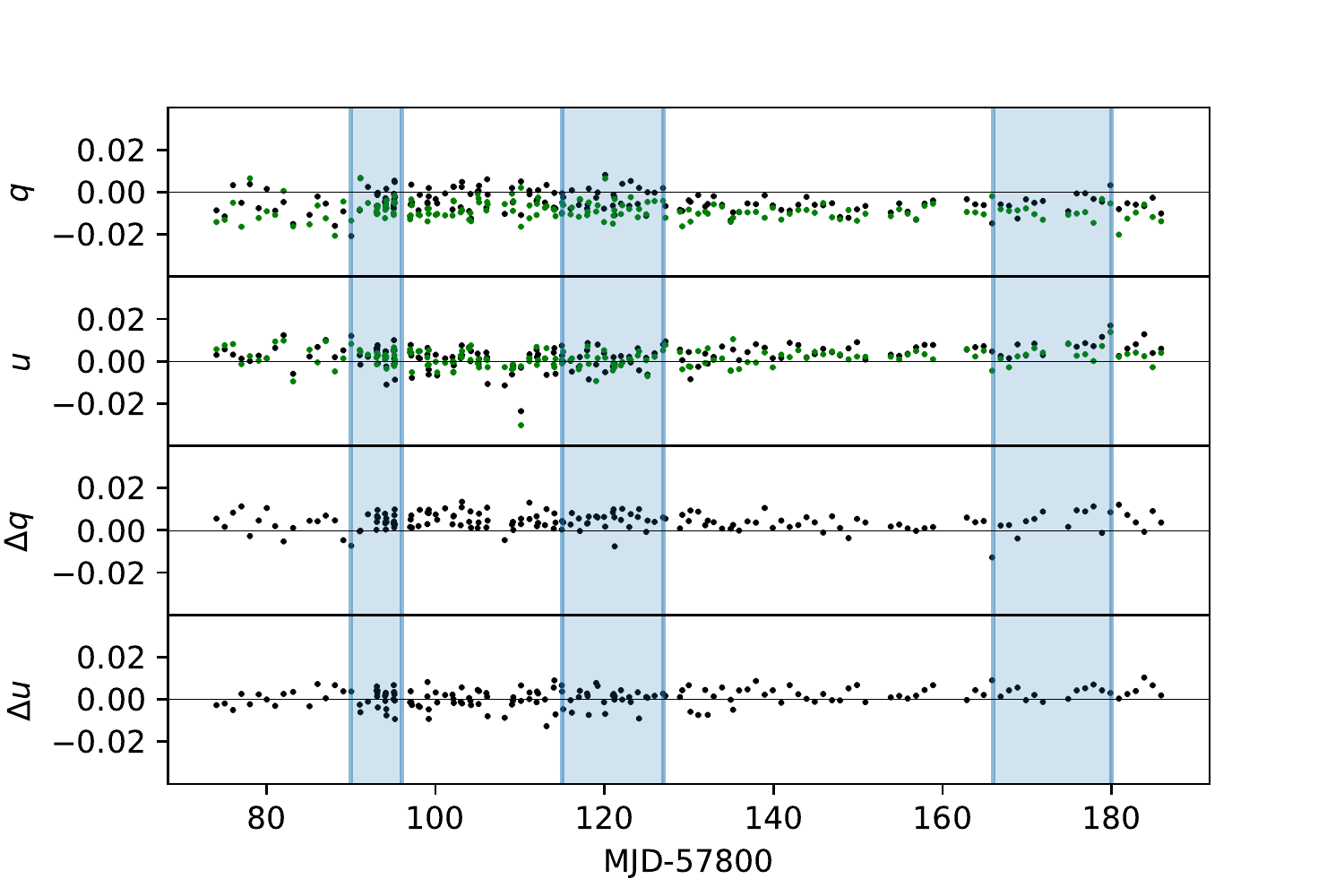}
   \caption 
   { \label{fig:red} $r^*$ ($\lambda~_{\rm eff}\sim 8500$-{\AA}) polarimetry versus date.  The upper two panels show the absolute measurements of the normalized Stokes parameters $q$ and $u$ for KIC 8462852 (black points) and the comparison object TYC-3162-977-1 (green points).  The Stokes parameters can be seen to partially track each other.  The lower two panels show the difference between the normalized Stokes parameters ($\Delta q$, $\Delta u$) of the two objects and therefore give relative polarimetry.  The vertical (blue) shaded regions are "dip" periods.}     
   \end{figure*} 

   \begin{figure*}
   \includegraphics[width=16.0cm]{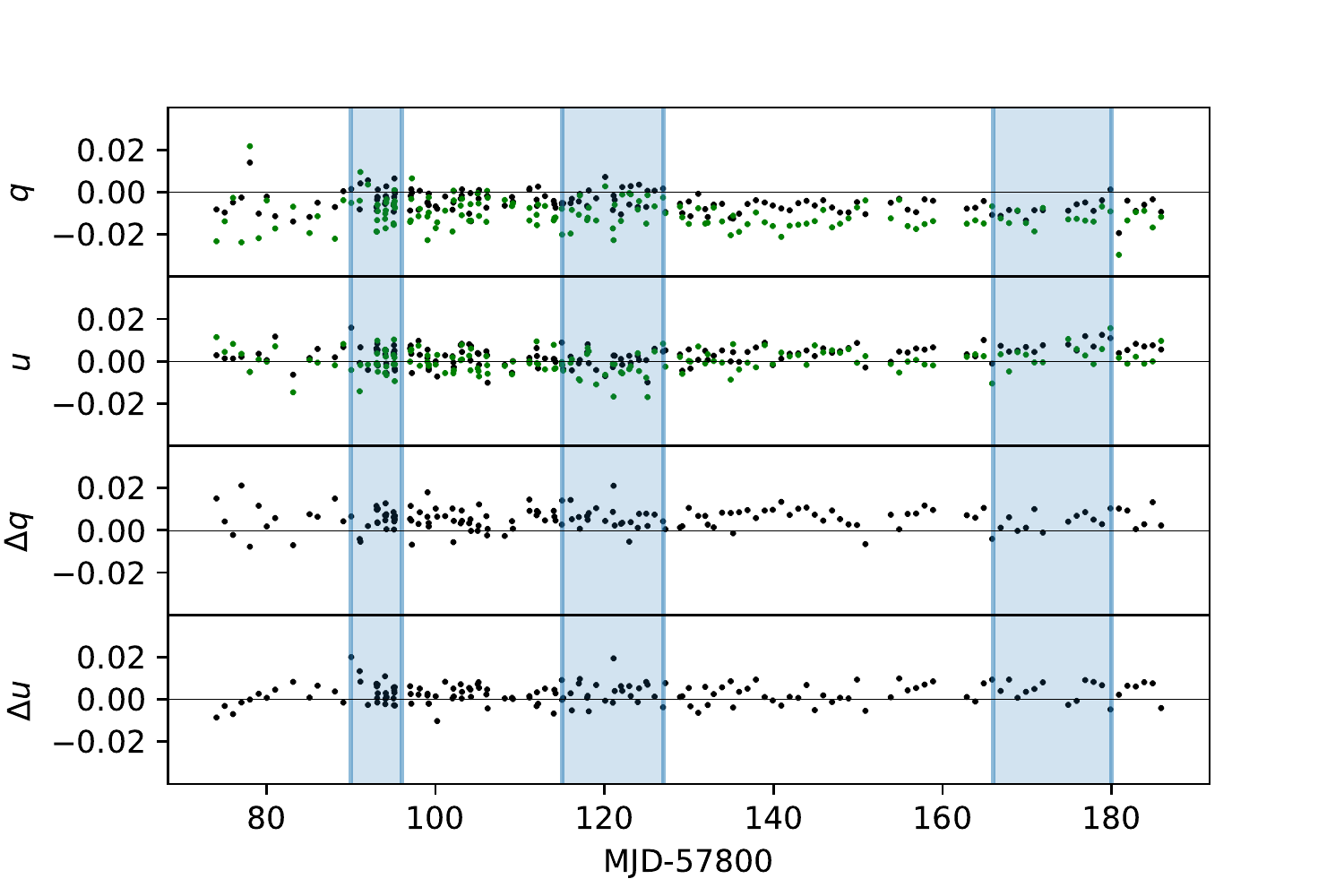}
   \caption 
   { \label{fig:green} $g^*$ ($\lambda~_{\rm eff}\sim 7050$-{\AA}) polarimetry versus date.  Other details as per Fig. \ref{fig:red}}
   \end{figure*} 

   \begin{figure*}
   \includegraphics[width=16.0cm]{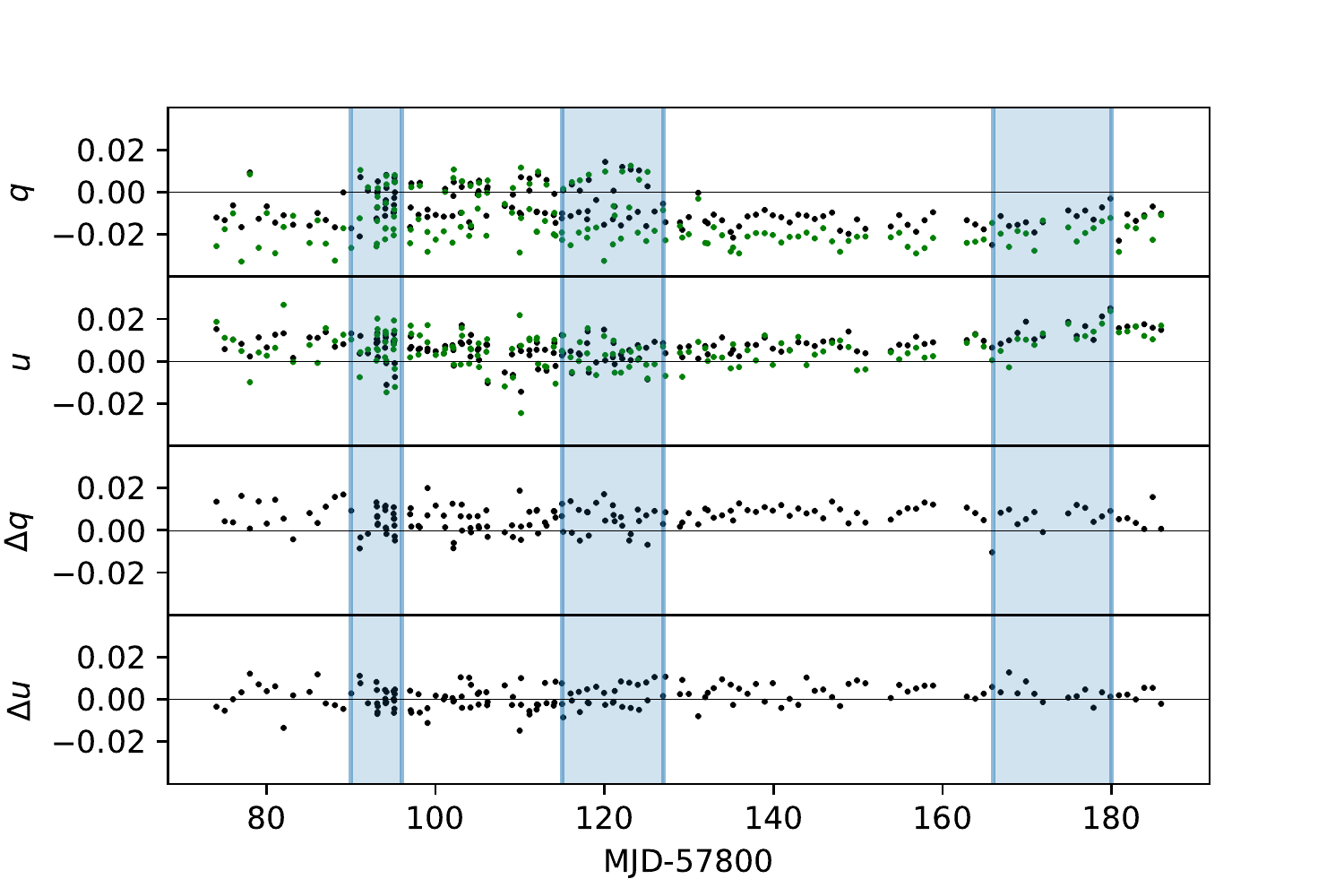}
   \caption 
   { \label{fig:blue} $b^*$ ($\lambda~_{\rm eff}\sim 5300$-{\AA}) polarimetry versus date. Other details as per Fig. \ref{fig:red}} 
   \end{figure*} 

\begin{figure}
\begin{center}
\includegraphics[width=8cm]{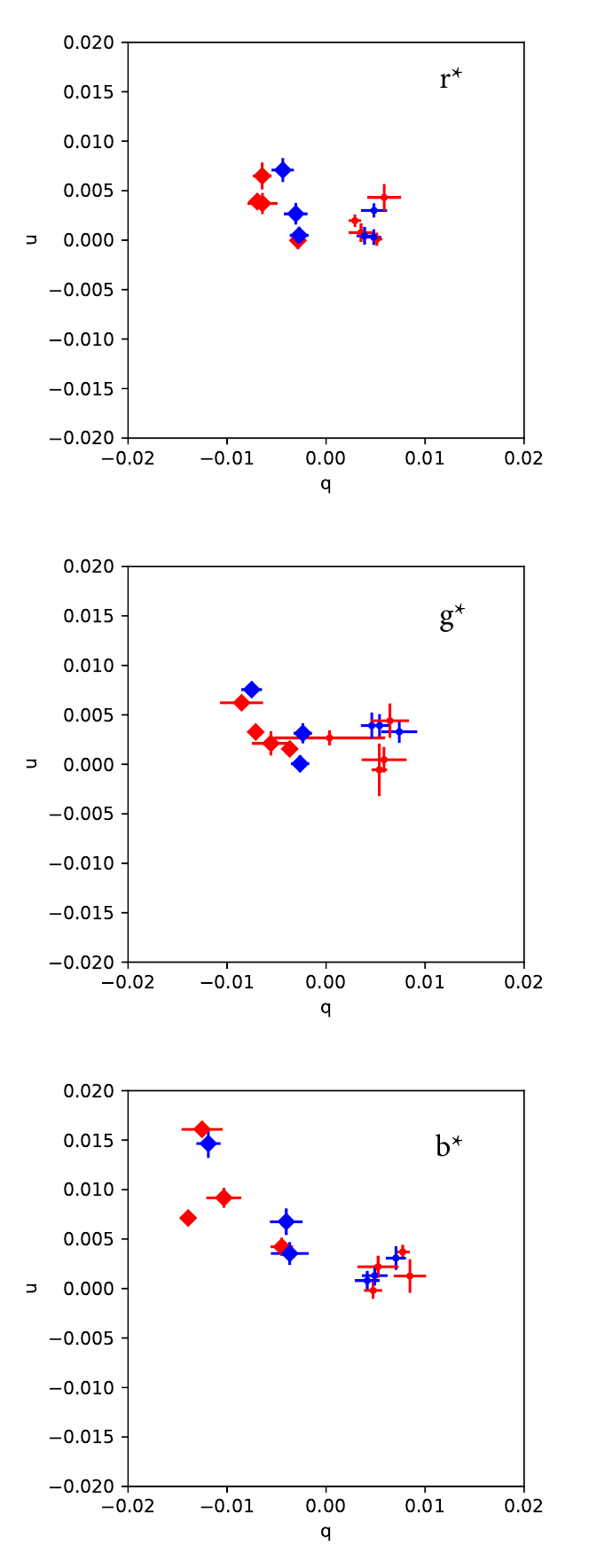}
\end{center}
\caption{\label{fig:qu} Stokes parameters in- (blue) and out- (red) of dip for the three wavebands.  Large diamond symbols indicate absolute Stokes parameters ($q,u$).  Small point symbols indicate relative ($\Delta q, \Delta u$) Stokes parameters.  The scatter of the relative Stokes measurements is reduced with respect to the absolute values.  In all cases the in- and out- of dip distributions overlap with no evidence of an associated change of polarization properties.}
\end{figure}

\section{Observations and Data Reduction}
\label{sec:obs}

As part of a longer term polarimetric monitoring campaign we observed the $4\times4$\arcmin field centred on KIC 8462582 a total of 166 times between 2017 April 30 and 2017 August 21. The observations were made with the RINGO3 \citep{ringo3} optical polarimeter of the Liverpool Telescope \citep{steele04}.  During this observing period three distinct dips were discovered by \cite{boy-atel,boy-prep}.  Inspection of their lightcurves\footnote{http://www.wherestheflux.com/} shows dips of $1-2$\% depth occurred during the periods MJD=57890-57896 (``Elsie''), 57915-57927 (``Celeste'') and 57968-57980 (``Skara Brae'').  In total 61 RINGO3 observations were made in-dip and 105 out-of-dip.

RINGO3 uses a rapidly rotating polarizer to modulate the incoming beam.  The modulated beam is then fed via a system of dichroic mirrors to three electron-multiplying CCD cameras.  These record the resulting time variable signal in eight equal-length exposures per camera per rotation. The dichroic mirrors define three wavebands we refer to as $b^*$ ($\sim 3500-6400$ {\AA}), $g^*$ ($\sim 6500 - 7600$ {\AA}) and $r^*$  ($\sim7700-10000$ {\AA}).  Convolving the dichroic transmission profiles with the detector quantum efficiency we find typical effective wavelengths for the three bands
of $\lambda_{\rm eff}\sim$ 5300, 7050 and 8500-{\AA}.

A detailed description of RINGO3 data reduction procedures is given by \citet{jermak-phd}; we therefore only give a brief summary here. After stacking the repeated exposures at the same rotation angle, bias and flat field corrections were applied.  We then used {\sc Sextractor} \citep{sextractor} to measure photometry of every object in each image.  We used a fixed aperture of 10 pixels (4.4 arcsec) diameter and local sky subtraction.  The 8 measured count values per object (one per rotation angle) per observation were then used to derive the instrumental Stokes $q$ and $u$ parameters for that object using the equations presented by \citet{clarke}.  
These were then corrected for the effect of instrumental polarization following the procedure outlined for RINGO2 in \citet{steele-grb}.  This correction used the mean $q$ and $u$ values of a set of calibration observations of non-polarized standard stars \citep{schmidt} obtained every $\sim5$ nights during the observing period.  Following this the measured q and u parameters were corrected for the effect of field rotation to derive the final $q$ and $u$ values in the sky plane.

\section{Results}
\label{sec:results}

\subsection{Absolute Polarimetry\label{section:abs}}
The upper two panels in Figures \ref{fig:red}, \ref{fig:green} and \ref{fig:blue} show our final Stokes $q$ and $u$ parameters for the three wavelength bands versus Modified Julian Date (MJD).
Making an initial assumption of no variability, the typical error (not plotted for claity) on each $q$ and $u$ point can be estimated from the standard deviation of the entire dataset.  We find standard deviations in $q$ and $u$ of 0.005 ($r^*$), 0.005 ($g^*$) and 0.007 ($b^*$).  Similar values were also found for two other objects in the field (TYC-3162-977-1 and TYC-3162-677-1). 
The standard deviations are much larger than the error calculated by photon counting statistics for each point (which are $<0.001$ for all points).  This is evidence of a previously known time variable systematic error that the instrument displays in all measurements \citep{slow}.

We split the reduced polarimetry curves into sections based on whether the source was dipping at the time.  This gives 4 out-of-dip sections and 3 in-dip.  Within each section we then calculated the mean and standard error of the mean of
$q$ and $u$.  These values are plotted (diamond symbols) in Figure \ref{fig:qu}.  It can be seen the standard errors
are considerably smaller than the spread between the mean values, indicating that long term variability is present.  The 
$q,u$ standard deviations of the distributions are 
0.0017,0.0025 ($r^*$), 0.0023, 0.0024 ($g^*$) and 0.0041,0.0045 ($b^*$).  However, given the fully overlapping distribution of the in- and out-of-dip points in Fig. \ref{fig:qu}, we attribute this variability to the systematic errors mentioned above rather than true variability of the source associated with the dips.  To obtain the final mean Stokes parameters of the source we therefore average the seven mean in and out of dip values, again calculating the standard error of the mean.  This yields 
$q,u= -0.0047\pm0.0007, 0.0035\pm0.0010$ ($r^*$), 
$q,u= -0.0053\pm0.0009, 0.0034\pm0.0011$ ($g^*$),
$q,u= -0.0087\pm0.0016, 0.0088\pm0.0018$ ($b^*$).

We then convert our final mean Stokes parameters to a mean degree of polarization ($P$) and associated error.  In doing this we correct for polarization bias and calculate error bars using a Monte Carlo implementation of the method of \cite{simmons}.
Overall we find averaged over the whole observation period the mean absolute polarization of the source is $P=0.6\pm0.1$\% ($r^*$), $P=0.6\pm0.1$\% ($g^*$) and $P=1.2\pm0.2$ ($b^*$)\% with corresponding electric vector polarization angles (EVPA) of $72\pm6^\circ$, $74\pm6^\circ$ and $73\pm6^\circ$.

For comparison we identified one other object that was within the RINGO3 field of view of all of our frames.  This object (TYC-3162-977-1) has $B=13.03, V=11.86$ and distance $d=826^{+602}_{-245}$ pc.  This is a similar apparent magnitude although with slightly redder colour and a greater distance than KIC 8462852 ($B=12.82, V=12.01, d=391^{+122}_{-75}$ pc).
The apparent magnitudes of both objects are quoted from the Tycho catalogue \citep{tycho} and the distances from Gaia Data Release 1 \citep{dr1}.  Applying the same techniques as outlined above, we measure polarizations for TYC-3162-977-1 of 
$P=0.9\pm0.1$\% ($r^*$), $P=1.0\pm0.1$\% ($g^*$) and $P=1.6\pm0.2$\% ($b^*$) with EVPA of $84\pm2^\circ$, $87\pm4^\circ$ and $77\pm3^\circ$ respectively.  The two sources therefore share comparable polarization characteristics, with $\sim 1.5 \times$ greater polarization for the source which is $\sim 2\times$ distant, EVPA within $\sim10^\circ$ of each other, and a similar dependence on wavelength.  We conclude that on average KIC 8462852 shows normal polarization properties consistent with the expected interstellar polarization for a source at its distance and location.  In particular there is no evidence of excess polarization.

\subsection{Relative Polarimetry}

In Figures \ref{fig:red}, \ref{fig:green} and \ref{fig:blue} (upper panels) we also plot our measurements of
TYC-3162-977-1 (green symbols).
The measurements of KIC 8462852 and TYC-3162-977-1 partially track each other, implying that the systematic errors identified in the previous section occur (at least partly) between observations rather than at different field positions within an individual observation.  We therefore calculate the difference between the Stokes parameters ($\Delta q$ and $\Delta u$) of the two objects.  In other words we treat TYC-3162-977-1 as a local polarimetric standard.  The results of this analysis are presented in the lower panels of Figures \ref{fig:red}, \ref{fig:green}
and \ref{fig:blue}.

Figure \ref{fig:qu} (point symbols) shows that this technique is effective in reducing the scatter on the resulting Stokes parameters, with the standard errors now similar in size to the scatter.  The standard deviation derived from the distribution of the mean $\Delta q$ and $\Delta u$ parameters are 0.0009,0.0014 ($r^*$), 
0.0020, 0.0018 ($g^*$) and 0.0006,0.0004 ($b^*$).  The mean improvement in standard deviation is factor 1.9 ($r^*$), 1.2 ($g^*$) and 9 ($b^*$) compared to
the standard absolute polarimetric analysis. This demonstrates the value of our relative technique in reducing the time variable systematic error.

Comparing the distribution of points in- and out-of-dip we again find complete overlap and no evidence of intrinsic variability 
in the polarization.  To quantify this we first calculate the difference between the mean Stokes parameters in- and out-of-dip 
(and associated errors via standard propagation theory).  We then convert to bias corrected polarization and error as per 
Section \ref{section:abs} to calculate $\delta P$, the difference between the in- and out-of-dip polarization.  We 
find $\delta P = 0.05\pm0.04$\% ($r^*$), $0.17\pm0.07$\% ($g^*$), $0.10\pm0.08$\% ($b^*$).  We note that
the error on these quantities should not be considered a measure of significance since for small 
signal-to-noise ratios errors on $P$ will be strongly non-Gaussian  due to its always positive nature \citep{wardle}.  Taking a conservative approach 
we therefore estimate any change in the degree of polarization between in- and out-of-dip states as upper limits of $<0.1$\% in the $r^*$ 
band and $<0.2$\% in the $g^*$ and $b^*$ bands.

\section{Discussion and Conclusions}
\label{sec:conculsions}

We have shown that the average polarization of KIC 8462852 is as
expected for its distance and location on the sky.  In addition we find a limit on any polarization change of $<0.1-0.2$\% between its dipping and non-dipping states. 

How these limits are interpreted depends on the specific model being considered.  If the dimming were due to 
a change in interstellar absorption along the line of sight to the star, a change in extinction of 1-2\% would be expected to 
yield a maximum polarization change $<0.05$\% \citep{serkowski}.   Alternatively if the occulting material were optically thick 
and thus covered only $\sim$1\% of the star the only polarimetric signal during the dimming events would arise from differences 
in polarisation across the stellar surface at levels significantly below our sensitivity \citep{carciofi}. At the other extreme, 
if the whole stellar disk were covered by material with an optical depth of $\sim$0.01 then the fractional polarisation introduced by 
this material must be less than 10-20\%.  Polarisation from dust-scattered light depends strongly on the scattering (phase) angle.  At 
angles of $\sim90^\circ$ levels of $P\sim15-30$\% are seen for Solar System comets \citep{comet-pol}, and up to $P\sim50$\% for 
circumstellar disks \citep{perrin}.  However $P$ drops to only a few percent for small angles, as is the case for the geometry of material transiting KIC~8462852. Thus our limits do not appear particularly constraining.

However, to illustrate the model-dependence we consider the comet family scenario proposed by \citet{boyajian} and explored in detail by \citet{wyatt}. In this scenario the dips are explained by a series of comet fragments and associated dust concentrations on highly eccentric orbits, which are viewed with their pericenter direction roughly along our line of sight. Thus, while the polarisation is expected to be low during a dimming event because the scattering angle is near zero, the highly eccentric orbit means that a few tens of days before or after the event the same dust concentration will pass through a point where the scattering angle is 90$^\circ$ (i.e. a true anomaly $f=90^\circ$). \citet{wyatt} show that the pericenter must be closer than 0.6-AU, at which point the time taken for a particle on a parabolic orbit to travel from $f=0$ to $f=90^\circ$ is about 50d.  As an illustration we calculate that if a Solar radius worth of dust area were present at 0.1-AU from KIC 8462852 and near a scattering angle of 90$^\circ$, and that dust had an albedo of 0.1 and polarisation fraction of 10\%, the polarization would be at the level of the observed RINGO3 sensitivity.  Thus, while our non-detection does not (yet) constrain the comet scenario, it predicts that even modest amounts of dust with properties similar to Solar System comets may be detectable.

Finally we note that the sensitivity of our measurements scales in inverse proportion to the depth of the dip/amount of material present.  For example detection of a future 0.2 magnitude dip would provide polarimetric sensitivity at the $0.5-1.0$\% level for the properties of any occulting dust.

\section*{Acknowledgements}

LT is operated on the island of La Palma by
Liverpool John 
Moores University in the Spanish Observatorio del Roque de los
Muchachos of the 
Instituto de Astrof\'{i}sica de Canarias with financial support from the 
UK STFC.  GMK is supported by the Royal Society as a University Research Fellow.

\bibliographystyle{mnras}
\bibliography{extra}





\bsp	
\label{lastpage}
\end{document}